\documentclass{jmlr}

\usepackage{longtable}
\usepackage{cleveref}

\usepackage[utf8]{inputenc}
\usepackage{amsfonts}
\usepackage{amssymb}
\usepackage{mathtools}
\usepackage{xspace}
\usepackage{paralist}
\usepackage{algorithm}
\usepackage[commentColor=blue,rightComments=false]{algpseudocodex}

\usepackage{footmisc}
\usepackage{color}
\usepackage{tikz}
\usetikzlibrary{arrows, automata, positioning, fit, shapes}

\tikzset{%
  sh2n/.style={shift={(0,0.31)}},
  sh2s/.style={shift={(0,-0.31)}},
  sh2e/.style={shift={(0.31,0)}},
  sh2w/.style={shift={(-0.31,0)}},
  sh2nw/.style={shift={(-0.31,0.31)}},
  sh2ne/.style={shift={(0.31,0.31)}},
  sh2sw/.style={shift={(-0.31,-0.31)}},
  sh2se/.style={shift={(0.31,-0.31)}},
  rc/.style={rounded corners=2mm},
  reflexive above/.style={->,loop,looseness=7,in=60,out=120, above},
  reflexive below/.style={->,loop,looseness=7,in=240,out=300, below},
  reflexive left/.style={->,loop,looseness=7,in=150,out=210, left},
  reflexive right/.style={->,loop,looseness=7,out=30,in=330, right},
}

\graphicspath{{figures/}}

\usepackage{graphics}

\usepackage[draft,layout=inline]{fixme}
\FXRegisterAuthor{rg}{arg}{RG}
\FXRegisterAuthor{as}{aas}{AS}
\FXRegisterAuthor{mf}{amf}{MF}
\FXRegisterAuthor{nw}{anw}{NW}

\usepackage{todonotes}

\newcommand{\hW}{\textit{hW}-inference\xspace}
\newcommand{\ehW}{\textit{ehW}-inference\xspace}

\newcommand{\splitatcommas}[1]{%
  \begingroup
  \begingroup\lccode`~=`, \lowercase{\endgroup
    \edef~{\mathchar\the\mathcode`, \penalty0 \noexpand\hspace{0pt plus 1em}}%
  }\mathcode`,="8000 #1%
  \endgroup
}

\jmlrvolume{}
\jmlryear{2024}
\jmlrworkshop{Learning and Automata (LearnAut) -- ICALP 2024 workshop}

\title[Learning EFSM]{Learning EFSM Models with Registers in Guards}

\author{
    \Name{Germán Vega} \Email{german.vega@imag.fr} \\
    \Name{Roland Groz} \Email{Roland.Groz@univ-grenoble-alpes.fr}\\
    \Name{Catherine Oriat} \Email{Catherine.Oriat@univ-grenoble-alpes.fr}\\
    \addr LIG, Universit\'{e} Grenoble Alpes, F-38058 Grenoble, France
    \AND
    \Name{Michael Foster} \Email{m.foster@sheffield.ac.uk} \\
    \Name{Neil Walkinshaw} \Email{n.walkinshaw@sheffield.ac.uk} \\
    \addr Department of Computer Science, The University of Sheffield, UK
    \AND
    \Name{Adenilso Sim\~{a}o} \Email{Adenilso@icmc.usp.br} \\
    \addr Universidade de S\~{a}o Paulo, ICMC, S\~{a}o Carlos/S\~{a}o Paulo, Brasil
}

\begin{document}

\maketitle

\begin{abstract}
This paper presents an active inference method for Extended Finite State Machines, where inputs and outputs are parametrized, and transitions can be conditioned by guards involving input parameters and internal variables called registers. The method applies to (software) systems that cannot be reset, so it learns an EFSM model of the system on a single trace.
\end{abstract}


\section{Introduction}\label{sec:introduction}


Automata learning by active inference has attracted interest in software engineering and validation since the turn of the century \cite{Peled99} \cite{Hagerer02}. Although classical automata (in particular Mealy models) have been used in software testing for a long time \cite{Chow78}, software engineering has evolved to richer models of automata, with parameters on actions, internal variables, and guards, as can be found in such formalisms as SDL, Statecharts or UML. Inferring such complex models, whose expressive power is usually Turing-complete, is challenging.

A significant step was the introduction of register automata \cite{isberner2014learning}. However, the practical application of those inference methods is limited by two main problems.
Firstly, the expressive power of the inferred models is usually restricted (e.g. only boolean values or infinite domains with only equality).
Secondly, most inference methods learn the System Under Learning (SUL) by submitting sequences of inputs from the \emph{initial state}, thus requiring the SUL to be reset for each query, which can be costly, impractical or impossible.

Recently, \cite{Foster2023ICFEM} proposed an algorithm to infer EFSMs in the above circumstances. It uses an approach based on the hW algorithm by \cite{groz2020hw} to infer the structure of the control state machine and to work around the absence of a reliable reset, and uses Genetic Programming \cite{DBLP:books/lulu/PoliLM2008}
to infer registers, guards and output functions from the corresponding data observations. This approach is, however, restrictive because guards on transitions can only involve input parameters and cannot be based on internal registers. Our challenge, therefore, is to provide a more general solution that not only infers the presence of registers and corresponding output functions but also enables us to include state transitions that depend upon register values.
\emph{Our paper proposes an algorithm that can infer EFSM models that include registers in guards.}

Addressing this problem raises a number of challenges.
\begin{inparaenum}[i)]
\item it is necessary to identify the underlying control state machine, by distinguishing what should (preferably) be encoded in states and what should be encoded in registers;
\item for any transition in the EFSM, it is necessary to identify any guards on inputs and the inferred registers, and to identify any functions that may affect the state of the registers or the values of the observable outputs; and
\item since no reliable reset is assumed, analyzing the influence of registers should be done from comparable configurations of the SUL, so it requires finding a way of coming back to a previously recognized configuration.
\end{inparaenum}

These challenges are tightly interdependent; any guards and output functions are highly dependent on the transition structure of the machine. Any inference process must generalise from the executions observed so far (the inferred model must be capable of recognising or accepting sequences that have not yet been observed). At the same time, it must avoid overgeneralization (i.e. accepting impossible or invalid sequences of inputs).

\section{Running example and background models}
We consider a straightforward running example that exhibits most features of the model yet is small enough to be fully illustrated in this paper. Our example is a vending machine, shown in \Cref{fig:drinks}, where the choice of drink and the money paid into the machine are parameters. These are recorded by registers (internal variables), that influence later computations. Throughout the paper, we refer to the parameterized events (such as $select(tea)$) as concrete inputs or outputs and to the event types (such as $select$) as abstract ones.

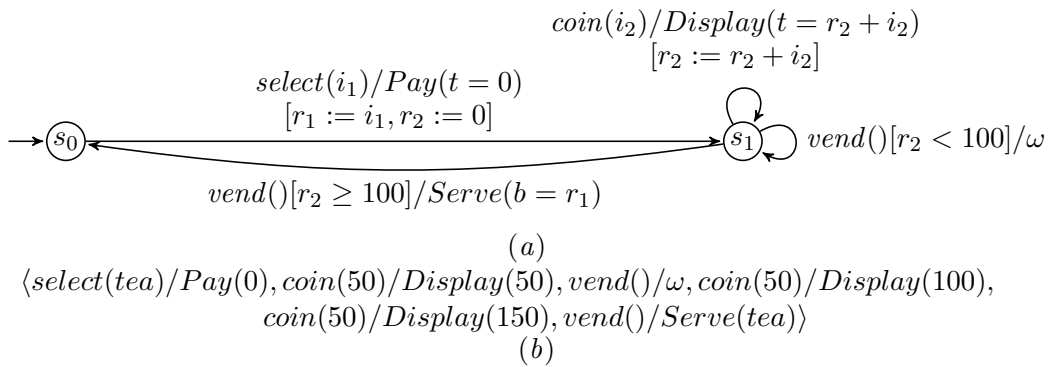
\begin{figure}[h!]
  \floatconts{fig:drinks}
  {\caption{Our vending machine EFSM and an example trace.}}
  {%
    \subfigure{%
      \label{fig:runningexample}
      \centering
      \begin{tikzpicture}[->,>=stealth',shorten >=1pt,auto, node distance=5cm, semithick, initial left, initial text={},
    state/.style={circle, draw, minimum size=0.3cm, inner sep=1}]
    \node[initial, state] (s0)               {$s_0$};
    \node[state, shift={(4, 0)}]          (s1) [right of=s0] {$s_1$};

    \draw (s0) edge [above]node[align=center, shift={(-0.2, 0.0)}]{$\mathit{select}(i_1)/Pay(t=0)$\\[-0.5mm]$[r_1:=i_1, r_2:=0]$}  (s1);
    \draw (s1) edge [reflexive above, above]node[align=center, shift={(-0.1, 0.1)}]{$\mathit{coin}(i_2)/Display(t=r_2+i_2)$\\[-0.5mm]$[r_2:=r_2+i_2]$} (s1);
    \draw (s1) edge [reflexive right]node[align=center, shift={(-0.0mm, 0.2mm)}]{$\mathit{vend}()[r_2 < 100]/\omega$} (s1);
    \draw (s1) edge [bend left=8, below]node{$\mathit{vend}()[r_2 \ge 100]/Serve(b=r_1)$}  (s0);
\end{tikzpicture}
    }
    \subfigure{%
      \label{fig:example-trace}
      \centering
      \parbox{\textwidth}{\centering $\langle select(tea)/Pay(0), coin(50)/Display(50), vend()/\omega, coin(50)/Display(100),$\newline
        $coin(50)/Display(150), vend()/Serve(tea) \rangle$}
    }
  }
\end{figure}

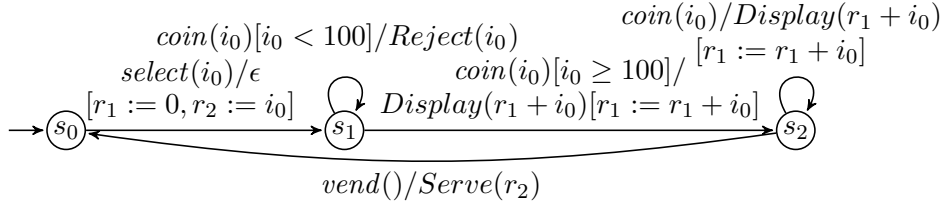
\begin{figure}[ht]
  \centering
  \begin{tikzpicture}[->,>=stealth',shorten >=1pt,auto, node distance=5cm, semithick, initial left, initial text={},
    state/.style={circle, draw, minimum size=0.3cm, inner sep=1}]
    \node[initial, state] (s0)               {$s_0$};
    \node[state, shift={(-1.3, 0)}]          (s1) [right of=s0] {$s_1$};
    \node[state, shift={(1, 0)}]          (s2) [right of=s1] {$s_2$};

    \draw (s0) edge [above]node[align=center, shift={(-0.2, 0.0)}]{$\mathit{select}(i_0)/\epsilon$\\[-0.5mm]$[r_1:=0, r_2:=i_0]$}  (s1);
    \draw (s1) edge [reflexive above, above]node[align=center, shift={(-0.1, 0.2)}]{$\mathit{coin}(i_0)[i_0 < 100]/Reject(i_0)$} (s1);
    \draw (s1) edge [sloped, above]node[align=center]{$\mathit{coin}(i_0)[i_0 \ge 100]/$\\[-0.3mm]$Display(r_1+i_0)[r_1 := r_1 + i_0]$} (s2);
    \draw (s2) edge [reflexive above]node[align=center, shift={(-0.2, 0.2mm)}]{$\mathit{coin}(i_0)/Display(r_1+i_0)$\\[-0.2mm]$[r_1 := r_1 + i_0]$} (s2);
    \draw (s2) edge [bend left=8, below]node{$\mathit{vend}()/Serve(r_2)$}  (s0);
\end{tikzpicture}
  \caption{No register allowed in guard (from \cite{Foster2023ICFEM}}
  \label{fig:icfem-efsm}
\end{figure}

\Cref{fig:icfem-efsm} is taken from \cite{Foster2023ICFEM}. In that work, guards can only be influenced by input parameters, not registers, so coins are rejected if they are less than the price of the drink. There are two distinct states after drink selection ($s_1$ and $s_2$), one where insufficient money has been input, and one where the drink can be served.
By allowing registers in guards, \Cref{fig:runningexample} represents a more realistic vending machine where we can pay with smaller coins until we reach a sufficient amount. A single state after selection is enough to allow different transitions depending on the amount stored in a register $r_2$.

The output $\omega$ of the $s_1 \xrightarrow{vend} s_1$ transition is a special output indicating that the transition neither produces visible output nor changes the system state. In a vending machine, this could model that the button cannot be pressed mechanically or is visibly ineffective.

Having named input and output parameters can also provide evidence that they share a common register value, as is the case for the $t$ output parameter in \Cref{fig:runningexample}, which represents the total amount of money put into the system by coins for a selected drink. $t$ is shared by output types $Pay$ and $Display$ and mapped to a single register $r_2$.

\section{Assumptions for tractability}\label{sec:assumptions}
In this work, an EFSM is a tuple $(Q,\mathcal{R},\mathcal{I},\mathcal{O},\mathcal{T})$ where
$Q$ is a finite set of states,
$\mathcal{R}$ is a cartesian product of domains, representing the type of registers.
$\mathcal{I}$ is the set of concrete inputs, structured as a finite set of abstract inputs $I$ each having associated parameters and their domains $P_I$. Similarly $\mathcal{O}$ for concrete outputs based on $O$ for abstract outputs.
$\mathcal{T}$ is a finite set of transition, i.e. tuples
$(s,x,y,G,F,U,s')$ where
$s, s' \in Q$, $x \in I$, $y \in O$,
$G: P_I(x) \times \mathcal{R} \rightarrow \mathbb{B}$ is the transition guard,
$F: P_I(x) \times \mathcal{R} \rightarrow P_O(y)$ is the output function that gives the value of the output parameters,
$U: P_I(x) \times \mathcal{R} \rightarrow \mathcal{R}$ is the update function that gives the value of the registers after the transition.

\begin{figure}[ht]
  \centering
  \begin{tikzpicture}[->,>=stealth',shorten >=1pt,auto, node distance=5cm, semithick, initial left, initial text={},
    state/.style={circle, draw, minimum size=0.3cm, inner sep=1}]
    \node[initial, state] (s0)               {$s_0$};
    \node[state, shift={(4, 0)}]          (s1) [right of=s0] {$s_1$};

    \draw (s0) edge [above]node[align=center, shift={(-0.2, 0.0)}]{$\mathit{select}/Pay$}  (s1);
    \draw (s1) edge [reflexive above, above]node[align=center, shift={(-0.1, 0.1)}]{$\mathit{coin}/Display$} (s1);
    \draw (s1) edge [reflexive right]node[align=center, shift={(-0.0mm, 0.2mm)}]{$\mathit{vend}/\omega$} (s1);
    \draw (s1) edge [bend left=8, below]node{$\mathit{vend}/Serve$}  (s0);
\end{tikzpicture}%
  \caption{Control NFSM of the vending machine.}
  \label{fig:vendingNFSM}
\end{figure}
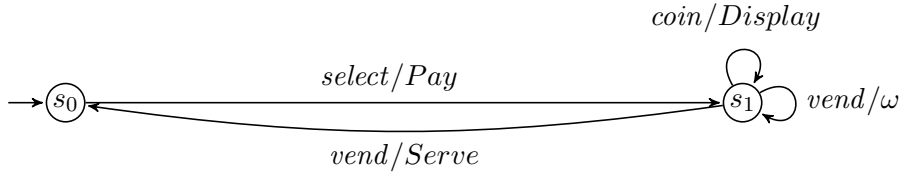

Associated with an EFSM, its \emph{control machine} is the non-deterministic finite state machine (NFSM) that results from abstracting from parameters (and therefore guards and registers). It defines the finite control structure of a given EFSM and is defined by a quadruple $(Q, I, O,\Delta)$, where $\Delta$ is the transition function
$\Delta: Q \times I \times O \to Q$, which lifted to sequences of inputs outputs as
$\Delta^*: Q \times (I \times O)^* \to Q$. The control machine for \Cref{fig:runningexample} is shown in \Cref{fig:vendingNFSM}.

We assume that the SUL is semantically equivalent to an EFSM, which has the following properties:
\begin{compactitem}
  \item It is \emph{deterministic} (at the concrete level), and its control NFSM is both \emph{strongly connected} (since we do not assume a reset, we can only learn a strongly connected component) and \emph{observable} (i.e. if an input triggers different transitions from the same state, they have different abstract outputs).
  \item Registers can only take values from input and output parameters, and are therefore observable: no hidden values influence the computation unnoticed. We further restrict them to storing only each parameter's last value (it could be extended to a bounded history). Note, however, that the guards and output functions can be arbitrarily complex, so this does not restrict the expressive power.
\end{compactitem}

Our method also relies on some inputs and hints provided on the system, which we assume can be given, although they might be approximate. The main goal is to reduce the practical complexity of the inference.
These will be shown on our example in \Cref{sec:example}. In particular, we assume we are given a characterization set $W \subset 2^{\mathcal{I}^+}$; and a homing sequence $h \in \mathcal{I}^*$, which also assigns a fixed value to all input parameters.

As the domains of input parameters can be infinite or very large (e.g. integers or floats), we will just use sample values in the learning process. We assume we are given (or pick) 3 levels of samples of concrete inputs $I_1 \subset I_2 \subset I_s$, with $I_1$ giving one concrete instance per abstract input, to infer the control machine, $I_2$ providing a few more concrete inputs to elicit guarded transitions, and $I_s$ being a larger set of samples to have enough data for the generalisation process to infer the output functions and guards.
No specific knowledge of the SUL is required to design those samples. We could pick either base values (e.g. for integers, 0 in $I_1$, plus 1 or -1 in $I_2$, and any further values in $I_s$) or just random values. The initial subsets will be extended when counterexamples provide new sample values that trigger so far unseen transitions.

We also assume we have 2 subsets of the registers $R_w \subseteq R_g \subseteq R$ such that $R_g$ are the only registers that can be used in guards, and $R_w$ are the registers used in guards that may be traversed when applying sequences from $W$ from any state. These can be overapproximated with $R_w = R_g = R$ in the worst case. However, if $R_w$ contains output registers, then the set of reachable values when applying inputs only from $I_1$ must be finite.

When we exercise a system, we learn \emph{instances} of transitions with concrete values for input and output parameters. For each transition of a control machine, we may have several sample values.
We collect these in a structure $\Lambda: Q \times I \times O \to 2^{\mathcal{R} \times \mathcal{I} \times \mathcal{O} \times \mathcal{R}}$, which associates abstract transitions from $\Delta$ with the observed concrete inputs and outputs and register configurations.

A \emph{sampled FSM}, which our backbone algorithm in \Cref{sec:example} will learn, is a quintuple $(Q,\mathcal{I},\mathcal{O},\Delta,\Lambda)$.
Our generalization process will then infer an EFSM from such a sampled machine.
The sampled machine for \Cref{fig:drinks} is shown in \Cref{fig:vendingNFSM}.

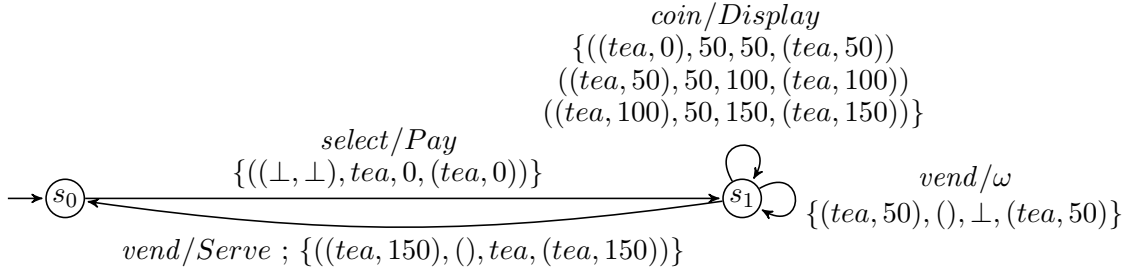
\begin{figure}[h]
  \centering
  \begin{tikzpicture}[->,>=stealth',shorten >=1pt,auto, node distance=5cm, semithick, initial left, initial text={},
    state/.style={circle, draw, minimum size=0.3cm, inner sep=1}]
    \node[initial, state] (s0)               {$s_0$};
    \node[state, shift={(4, 0)}]          (s1) [right of=s0] {$s_1$};

    \draw (s0) edge [above]node[align=center, shift={(-0.2, 0.0)}]{$\mathit{select}/Pay$\\[-0.5mm]$\{ ((\bot,\bot),tea,0,(tea,0))\}$}  (s1);
    \draw (s1) edge [reflexive above, above]node[align=center, shift={(-0.1, 0.1)}]{$\mathit{coin}/Display$\\
      [-0.5mm]$\{ ((tea,0),50,50,(tea,50))$\\
      [-0.5mm]$ ((tea,50),50,100,(tea,100))$\\
      [-0.5mm]$ ((tea,100),50,150,(tea,150))\}$}
    (s1);
    \draw (s1) edge [reflexive right]node[align=center, shift={(-0.0mm, 0.2mm)}]{$\mathit{vend}/\omega$\\
    $\{(tea,50),(),\bot,(tea,50)\}$} (s1);
    \draw (s1) edge [bend left=8, below]node{$\mathit{vend}/Serve$ ; $\{ ((tea,150),(),tea,(tea,150))\}$}  (s0);
\end{tikzpicture}
  \caption{Sampled FSM with trace from \Cref{fig:example-trace}}
  \label{fig:vendingsampled}
\end{figure}

Finally, we assume we are in the MAT framework from \cite{angluin1988queries}, where an oracle can provide counterexamples: this can easily be approximated by a random walk on the inferred machine (NFSM or EFSM) where (abstract, resp. concrete) outputs can be compared with the observed responses from the SUL.

\section{Execution of method on running example}
\label{sec:example}

At its core, our \ehW{} algorithm (see Annex) adapts the Mealy \hW{} algorithm by \cite{groz2020hw}, the so-called backbone, to learn the control structure of the EFSM on abstract inputs and outputs. The fact that registers can influence guards has two implications. First, from a given state, the same concrete input can produce two different abstract outputs. Second, and more importantly, characterisations (abstract responses to $W$) may be influenced by the register configuration (at least registers from $R_w$). 

Therefore during the learning process, we can only conservatively infer that we are in the same state if we observe the same characterisation \emph{from the same register configuration}. So in this new algorithm, the state space of the inferred NFSM is : $Q \subset 2^{W \to O^+} \times \mathcal{R}_w$. Another difference is that to be able to incrementally characterise a state (or learn a transition from it) we need to reach it with the same register configuration. For this reason, we require a stronger notion of homing that always resets the registers, and there is additional complexity for transferring to the next state or transition to learn (as reflected in \Cref{func:transfer}).

In our example from \Cref{fig:runningexample}, there are four observable input/output parameters ($i_1$ for choice of drink, $i_2$ for the value of the coin, $t$ for the total amount already inserted, and $b$ for the drink served). So there are at most 4 registers needed (let us name them $i_1,i_2,b,t$), and the algorithm will track the last value of each.

We pick $W=\{\textit{select}(\textit{coffee})\}$. The abstract output sequence $\textit{Pay}$ will characterise $s_0$ whereas the sequence $\Omega$ (inapplicable input) characterises $s_1$. The only guard in our example is on input $vend$, so our $W$ will never traverse it. But to show the robustness of the approach, let us assume we do not have this precise information, and think the choice of drink could play a role (which might be the case if drinks had different prices). So we pick $R_w=\{i_1\}$.
As for other guards, let us assume we have no clue, so we pick $R_g=R=\{i_1,i_2,b,t\}$.

We pick as register homing sequence $h=\textit{coin}(100).\textit{vend}.\textit{select}(\textit{coffee})$. Notice that for this simple example there are other shorter homing sequence, but we need one that resets the registers from $R_g$ (the chosen $h$ will reset values of $i_1$ to $\textit{coffee}$, $i_2$ to $100$ and $t$ to $0$). Notice also that this is actually a resetting sequence to $s_1$, but this may not be the case for more complex examples.

Finally, we pick $I_1=\{\textit{coin}(100),\textit{select}(\textit{coffee}),\textit{vend}\}$.
We always pick $I_1$ such that it includes all concrete inputs from $h$ and $W$, so as to be able to follow transitions when walking the graph of the NFSM to transfer to the next transition to learn.
And $I_2=I_s=I_1 \cup \{\textit{coin}(50),\textit{coin}(200),\textit{select}(\textit{tea})\}$.


\newcommand{\event}[2]{%
	\begin{array}[b]{c}
		{\scriptstyle #1}\\ [-1ex]
	   \rightarrow {_{\scriptstyle #2}}
	\end{array}
}

\newcommand{\Q}[2]{%
	#2,\register{#1}
}

\newcommand{\register}[1]{%
	\langle#1\rangle
}

\newcommand{\state}[2][?]{%
	\begin{array}[c]{c}
		{\scriptstyle #2}\\
		\hline {\scriptstyle #1}
	\end{array}
}

\newcommand{\step}[2]{\underbrace{#2}_{#1}}
\newcommand{\h}[2][h]{\step{#1}{#2}}
\newcommand{\W}[2][W]{\step{#1}{#2}}
\newcommand{\transfer}[2][transfer]{\step{#1}{#2}}
\newcommand{\X}[2][X]{\step{#1}{#2}}
\newcommand{\s}[2][s]{\step{#1}{#2}}
\newcommand{\CE}[2][CE]{\step{#1}{#2}}

\begin{figure}[ht]
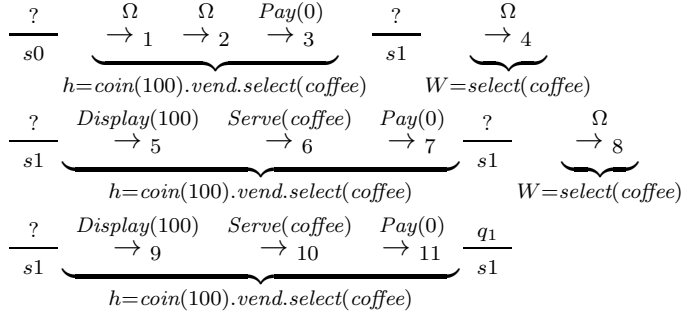

	\begin{tabular}[c]{l}
	$
	\state[s0]{?}
	\h[h=\textit{coin}(100).\textit{vend}.\textit{select}(\textit{coffee})] {
	\event{\Omega}{1}
	\event{\Omega}{2}
	\event{\textit{Pay}(0)}{3}
	}
	\state[s1]{?}
	\W[W=\textit{select}(\textit{coffee})] {
	\event{\Omega}{4}
	}
	$
	\\
	$
	\state[s1]{?}
	\h[h=\textit{coin}(100).\textit{vend}.\textit{select}(\textit{coffee})] {
	\event{\textit{Display}(100)}{5}
	\event{\textit{Serve}(\textit{coffee})}{6}
	\event{\textit{Pay}(0)}{7}
	}
	\state[s1]{?}
	\W[W=\textit{select}(\textit{coffee})] {
		\event{\Omega}{8}
	}
	$
	\\
	$
	\state[s1]{?}
	\h[h=\textit{coin}(100).\textit{vend}.\textit{select}(\textit{coffee})] {
	\event{\textit{Display}(100)}{9}
	\event{\textit{Serve}(\textit{coffee})}{10}
	\event{\textit{Pay}(0)}{11}
	}
	\state[s1]{q_1}
	$
	\end{tabular}
	\caption{Homing tails characterisations (internal SUL states shown only for reference) }
    \label{fig:trace-a}
\end{figure}

\Cref{fig:trace-a} shows how we first learn the homing tails. In steps 1-4 (steps are denoted $\rightarrow n$ in \Cref{fig:trace-a}), we learn that $H(\Omega.\Omega.\textit{Pay})=q_1=(\Omega,\textit{coffee})$, i.e. that after seeing the abstract output sequence $\Omega.\Omega.Pay$ the state we reach is characterised by the abstract output sequence $\Omega$ when $W$ is executed in $R_w$ configuration $i_1=\textit{coffee}$. Similarly, in steps 5-8 we learn that $H(\textit{Display}.\textit{Serve}.\textit{Pay})=(\Omega,\textit{coffee})=q_1$. After a new homing (steps 9-11), we observe a previously recorded output sequence, so we know the current state ($q_1$) and we can start learning its transitions.

\begin{figure}[ht]
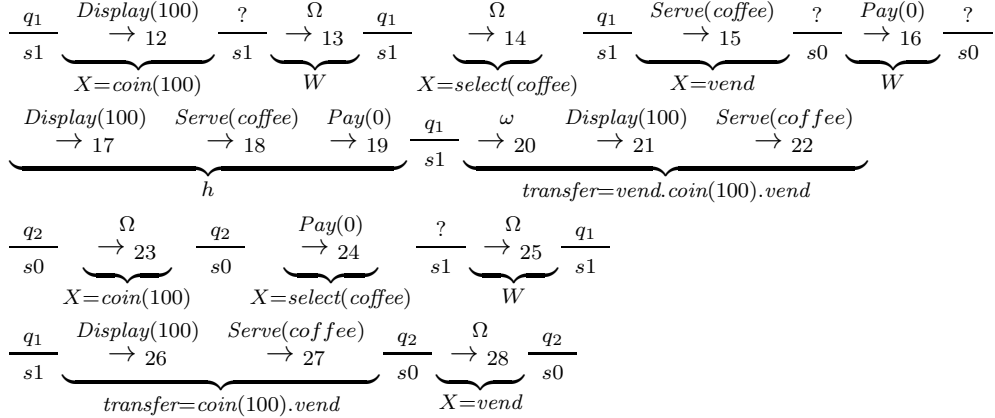

	\begin{tabular}[c]{l}
	$
	\state[s1]{q_1}
	\X[X={\textit{coin}(100)}]{
	\event{\textit{Display}(100)}{12}
	}
	\state[s1]{?}
	\W{
	\event{\Omega}{13}
	}
	\state[s1]{q_1}
	\X[X=\textit{select}(\textit{coffee})] {
	\event{\Omega}{14}
	}
	\state[s1]{q_1}
	\X[X=vend]{
	\event{\textit{Serve}(\textit{coffee})}{15}
	}
	\state[s0]{?}
	\W{
	\event{\textit{Pay}(0)}{16}
	}
	\state[s0]{?}
	$
	\\
	$
	\h{
	\event{\textit{Display}(100)}{17}
	\event{\textit{Serve}(\textit{coffee})}{18}
	\event{\textit{Pay}(0)}{19}
	}
	\state[s1]{q_1}
    \transfer[\textit{transfer}=\textit{vend}.\textit{coin}(100).\textit{vend}] {
	\event{\omega}{20}
	\event{\textit{Display}(100)}{21}
	\event{\textit{Serve}(coffee)}{22}
	}
	$
	\\
	$
	\state[s0]{q_2}
	\X[X=\textit{coin}(100)]{
	\event{\Omega}{23}
	}
	\state[s0]{q_2}
	\X[X=\textit{select}(\textit{coffee})] {
	\event{\textit{Pay}(0)}{24}
	}
	\state[s1]{?}
	\W{
	\event{\Omega}{25}
	}
	\state[s1]{q_1}
	$
	\\
	$
	\state[s1]{q_1}
	\transfer[\textit{transfer}=\textit{coin}(100).\textit{vend}] {
	\event{\textit{Display}(100)}{26}
	\event{\textit{Serve}(coffee)}{27}
	}
	\state[s0]{q_2}
	\X[X=\textit{vend}]{
	\event{\Omega}{28}
	}
	\state[s0]{q_2}
	$
	\end{tabular}
	\caption{NFSM learning (internal SUL states shown only for reference) }
    \label{fig:trace-b}
\end{figure}

For each learnt state, we try to infer a transition using each of the concrete inputs in $I_1$. We start from the current state $q_1$ with input $\textit{coin}(100)$, see steps 12-13 in \Cref{fig:trace-b}, and we discover a self loop. In step 14 we discover that input $\textit{select}$ is not allowed in the state, so we stay in the same state and can immediately learn the next input. Finally, in step 15 we discover that input $\textit{vend}$ leads to a newly discovered state $q_2=(\textit{Pay},\textit{coffee})$.
Thus $\Delta(q_1,\textit{vend},\textit{Serve})=q_2$, and
$\splitatcommas{\Lambda(q_1, \textit{vend}, \textit{Serve}) = ([100, \textit{coffee}, 100, \textit{coffee}], \textit{vend}, \textit{Serve}(\textit{coffee}), [100, \textit{coffee}, 100, \textit{coffee}])}$.
Note that after applying $W$ from $q_2$ we do not know the current state in the trace (shown as ``?''), so we need to home again.

At this point we try to come back to state $q_2$ to fully learn its transitions. First we home to a known state (steps 17-19 in \Cref{fig:trace-b}). This leads to $q_1$, from which we have learnt all inputs from $I_1$, so we follow a path in the partially inferred NFSM automaton to transfer (steps 20-22 in \Cref{fig:trace-b}) to $q_2$. Compared to Mealy learning, there is an additional complexity because we need to reach not just a node of the graph, but also a specific register configuration.
In step 20 in \Cref{fig:trace-b}, we optimistically try the shortest path (follow transition $vend$), but that fails because we observe a new abstract output $\omega$ for the same input. Thus we learnt that $vend$ from $q_1$ triggers a guarded transition with at least two possible abstract outputs. Since the new output is $\omega$ we know it did not change the state, so we are still in $q_1$ and need to transfer to $q_2$.
To reach that goal, we need to reach the register configuration $[100,\textit{coffee},100,\textit{coffee}]$ that we know enabled $\textit{vend}$. The transfer sequence of steps 21 and 22 does this.
After reaching $q_2$, we are able to learn the transitions for input $\textit{coin}(100)$ (step 23) and $\textit{select}(\textit{coffee})$ (step 24). Finally we need to do a new transfer to learn transition $\textit{vend}$ (step 28), and we have an NFSM that is complete with respect to $I_1$.

\begin{figure}[ht]
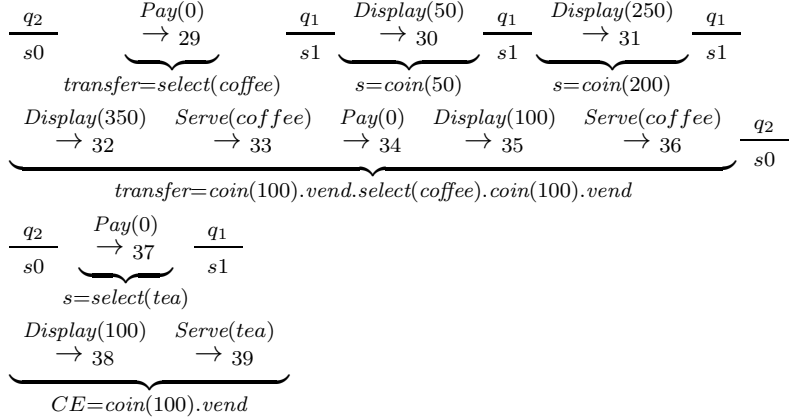

	\begin{tabular}[c]{l}
	$
	\state[s0]{q_2}
	\transfer[\textit{transfer}=\textit{select}(\textit{coffee})] {
		\event{\textit{Pay}(0)}{29}
	}
	\state[s1]{q_1}
	\s[s=\textit{coin}(50)] {
	\event{\textit{Display}(50)}{30}
	}
	\state[s1]{q_1}
	\s[s=\textit{coin}(200)]{
	\event{\textit{Display}(250)}{31}
	}
	\state[s1]{q_1}
	$
	\\
	$
	\transfer[\textit{transfer}=\textit{coin}(100).\textit{vend}.\textit{select}(\textit{coffee}).\textit{coin}(100).\textit{vend}] {
		\event{\textit{Display}(350)}{32}
		\event{\textit{Serve}(coffee)}{33}
		\event{\textit{Pay}(0)}{34}
		\event{\textit{Display}(100)}{35}
		\event{\textit{Serve}(coffee)}{36}
	}
	\state[s0]{q_2}
	$
	\\
	$
	\state[s0]{q_2}
	\s[s=\textit{select}(\textit{tea})]{
		\event{\textit{Pay}(0)}{37}
	}
	\state[s1]{q_1}
	$
	\\
	$
	\CE[CE=\textit{coin}(100).\textit{vend}] {
		\event{\textit{Display}(100)}{38}
		\event{\textit{Serve}(tea)}{39}
	}
	$
	\end{tabular}
	\caption{EFSM sampling (internal SUL states shown only for reference) }
    \label{fig:trace-c}
\end{figure}

Once we have found the control structure (NFSM), we walk the graph to try incrementally each input from $I_s=I_1 \cup \{\textit{coin}(50),\textit{coin}(200),\textit{select}(\textit{tea})\}$ in each transition, so as to collect enough samples for the \textsc{generalise} procedure to infer guards and output functions to build the EFSM. Notice that if we have previouosly learnt that an input is not allowed in a given state ($\Omega$ transitions), there is no need to sample further parameters for this transition.

\Cref{fig:trace-c} shows the sampling phase for our example, for state $(\Omega,\textit{coffee})$ we sample $\textit{coin}(50)$ in step 30 and $\textit{coin}(200)$ in step 31. For state $(\textit{Pay},\textit{coffee})$ we sample $\textit{select}(\textit{tea})$ in step 37. Notice that the most complex part is to compute the transfer sequence to reach the $R_g$ register configuration that enables a given transition. However, in the worst case, as the state was reached from some homing that resets all $R_g$ registers, there always exists a path from some homing that leads to this previously visited state and register configuration.

In more complex examples, after sampling it is possible to identify states that are equivalent in the NFSM structure and compatible with the collected  samples. These states can be merged in the EFSM, as we can infer, given the evidence, that they correspond to different register configurations of the same SUL state (this is done by the \textsc{reduceFSM} procedure).

Once the control structure (NFSM) is ascertained, we can use genetic programming to generalise from the collected samples to get the EFSM as was done in \cite{Foster2023ICFEM}. This is in fact simpler than in \cite{Foster2023ICFEM} because, as discussed in \Cref{sec:assumptions}, we know the values of the internal registers (and how they are updated) at each point in the trace. Thus, we only need to infer output functions and guards.

Notice in our example that even after sampling and generalisation, we have no evidence that the parameter of $\textit{Serve}$ can be anything other than $\textit{coffee}$. We rely on the oracle to provide further counterexamples (see steps 38-39) that can be used to refine the functions. After step 39, the \textsc{generalise} procedure produces exactly the EFSM of \Cref{fig:runningexample} (modulo renaming).


\section{A more complex example where $W$ traverses guards}
To illustrate the need for other key elements of our method, we highlight them on a slightly more complex example that entails problematic features.
\begin{figure}[ht!]
  \centering
  \begin{tikzpicture}[->,>=stealth',shorten >=1pt,auto,
node distance=5cm, semithick, initial left, initial text={},
    state/.style={circle, draw, minimum size=0.3cm, inner sep=1},
    ]
    \node[initial, state] (s0)               {$s_0$};
    \node[state]          (s1) [right of=s0, shift={(0, 1.5)}] {$s_1$};
    \node[state]          (s2) [right of=s1, shift={(0, -1.5)}] {$s_2$};
    \node[state]          (s3) [above of=s1, shift={(0, -3cm)}] {$s_3$};

    \draw[sloped, above, bend left=15] (s0) edge node[align=center]{$a(i_a/A(i_a+r_b)$}  (s1);
    \draw[sloped, above, bend left=35] (s0) edge node[align=center]{$b(i_b)/B$}  (s3);

    \draw[sloped, above, bend left=15] (s1) edge node[align=center]{$a(i_a)[r_a \ge r_b]/B$}  (s2);
    \draw[above] (s1) edge node[align=center, shift={(0, -0.5)}]{$a(i_a)[r_a < r_b]/$\\$A(i_a * r_b)$}  (s3);
    \draw[sloped, above, bend left=15] (s1) edge node[align=center]{$b(i_b)/B$}  (s0);

    \draw[sloped, above, bend left=15] (s2) edge node[align=center]{$a(i_a)/B$}  (s1);
    \draw[sloped, above] (s2) edge node[align=center]{$b(i_b)/B$}  (s0);

    \draw[sloped, above, bend right=15] (s3) edge node[align=center]{$a(i_a)/A(i_a \times r_b)$}  (s0);
    \draw[sloped, above, bend left=15] (s3) edge node[align=center]{$b(i_b)[i_b < r_a]/B$}  (s2);
    \draw (s3) edge [reflexive above, right]node[align=center, shift={(0.2, -0.2)}]{$b(i_b)[i_b \ge r_a]/B$} (s3);
\end{tikzpicture}
  \caption{$W=\{b(0).a(0)\}$ from state $s_3$ can yield $A.B$ or $B.A$}
  \label{fig:ades-bb-2}
\end{figure}
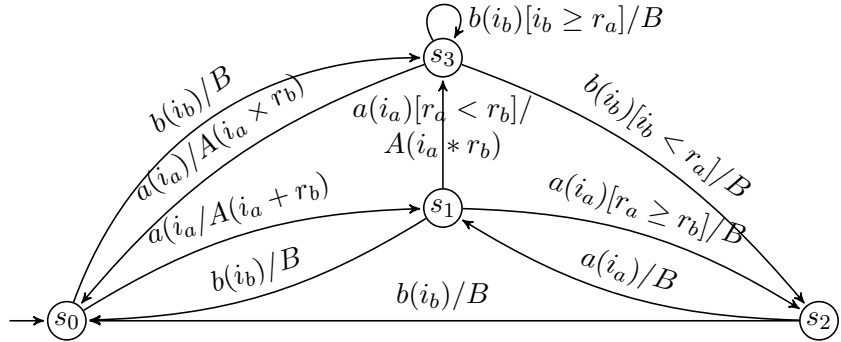
In \Cref{fig:ades-bb-2}, the provided $W=\{b(0).a(0)\}$ can have two outcomes when applied from state $s_3$: $A.B$ (if $r_a \leq 0$) or $B.A$ (otherwise); and $B.A$ is also the output for $W$ from state $s_1$. Actually, $\{b(0).a(0)\}$ is not even fully characterizing, but just as \hW{}, the algorithm is relatively robust and can infer the correct structure from approximately correct $h$ and $W$.
There are 3 registers linked to input and output parameters, which we name $r_a,r_b,r_A$. $R_g=\{r_a,r_b\}$. We overapproximate $R_w=\{r_a,r_b\}$ even though in fact $r_a$ is the only register really traversed when we know the SUL.
When we apply the algorithm with $I_1=\{a(0),b(0)\}$ and $I_2=I_1 \cup \{a(1),b(1)\}$ there will be different states in the NFSM corresponding to state $s_3$, namely $\splitatcommas{(A.B,[0,0]),(A.B,[0,1]),(B.A,[1,0]),(B.A,[1,1])}$. But we shall also have several NFSM states corresponding to a single other state, for instance $\splitatcommas{(B.B,[0,0]),(B.B,[1,0])}$ for state $s_0$. The backbone algorithm comes up with an 11-state NFSM, but the \textsc{reduceFSM} procedure would merge redundant (equivalent) copies so we can get a sampled machine with only 4 states at step 310. This number of steps was caused by our overapproximation on $R_w$ that led to exploring 11 states and 28 transitions.
The whole inference process reached 372 steps for that machine when we added sample values $\{a(-5),b(-5)\}$ to $I_s$ (before reducing) to get the exact data functions.

\section{Conclusion}
In this paper, we investigated how to infer EFSM models that include registers in guards. By allowing registers to be used in guards, the inference method presented here should be applicable to many systems. There are still a number of issues we want to investigate to consolidate it.
First, we assumed here that we could be given correct $h$ and $W$ for the SUL. But just as in \hW{}, we can look for inconsistencies that could reveal the need to expand $h$ or $W$, which could even be empty (not provided) at the start. Extending $h$ as in \hW{} is straightforward, but there are more options for $W$.
Another challenge is to find optimal transfer sequences. In our current preliminary implementation, we fall back on homing in all cases before transferring, and using a cached access table to states and configurations, but shorter transfers would reduce the length of the inference trace.
Finally, we would like to assess the method and its scalability. It is classical to work with randomly generated machines, but for EFSMs, there are many parameters to define what a random EFSM should be. Although there are benchmarks for finite automata\footnote{\url{http://automata.cs.ru.nl/Overview\#Mealybenchmarks}}, it is not easy to find strongly connected models with parameterized I/O behaviour. And automatically assessing that the inferred model is correct can also be challenging, since in the general case, equivalence of two EFSMs is undecidable.


\small
\let\oldbibitem=\bibitem
\def\bibitem{\vspace*{-6pt}\oldbibitem}
\bibliography{main}

\vspace{1cm}
\SetStartEndCondition{ }{}{}%
\SetKwFunction{Range}{range}
\SetKw{KwTo}{in}\SetKwFor{For}{for}{\string:}{}%
\SetKwIF{If}{ElseIf}{Else}{if}{ then}{elif}{else}{}%
\SetKwFor{While}{while}{:}{fintq}%
\SetKwRepeat{Repeat}{\addtocounter{AlgoLine}{-1}repeat}{\nl until}%
\SetKwFor{For}{for}{ do}{}



\AlgoDontDisplayBlockMarkers\SetAlgoNoEnd
\DontPrintSemicolon

\SetKwFunction{Backbone}{\textrm{\textsc{Backbone}}}
\SetKwFunction{Home}{\textrm{\textsc{Home}}}
\SetKwProg{Fn}{Function}{}{}
\SetKw{Init}{Initializing:}

\setcounter{AlgoLine}{0}

\begin{algorithm}
\nl        \Fn(\Comment{Apply h until we can know the state reached}){\Home{$T,r,h$}}{
\nl        \Repeat{$q \neq \bot$} {
\nl             $(T,a,r) \gets \textsc{Apply}(T,r,h)$ \Comment{Apply homing seq $h$, observe response $a$, update registers} \label{home:apply-homing}
\nl           let $\eta=\pi(a) \in O^*$
            \Comment{State reached by homing is associated to $\eta$. $\pi$ is abstraction}\\
\nl            \If(\Comment{Learn characterization of the tail state}){$H(\eta)$ is undefined for some $w \in W$}{  \label{home:partially-learnt}
\nl                $(T,y,r) \gets \textsc{Apply}(T,r,w)$\\
\nl                $H(\eta) \gets H(\eta) \cup \{w \mapsto \pi(y)\}$ \label{home:add-w}
}
\nl            \Else(\Comment{We know the state reached after $h/a$}){
\nl                let $q = (H(\eta),\rho_w(r,\epsilon))$ be the state reached at end of $h$
  \Comment{$\rho_w(r,\sigma)$ first applies register updates from i/o sequence $\sigma$ to $r$, then projects on $R_w$}\\
\nl                $Q \gets Q \cup \{q\}$; $A(\pi(a))(q,r) \gets \epsilon$
  \Comment{$A$ records known access to configuration $(q,r)$ from homing tail}
}
}
\nl         \Return $(T,q,r)$
}
    \caption{Homing into a known state}
    \label{func:home}
\end{algorithm} 

\setcounter{AlgoLine}{0}

\begin{algorithm}[h]
    \nl \Fn{\Backbone{$T,I_1,I_s,h,W,Q,\Delta,\Lambda$}}{
    \nl    \Init{$H \gets \emptyset$, $J \gets I_1$,
                 $q \gets \bot$, $r \gets \bot$} \\
    \nl    \Repeat{$\Delta,\Lambda$ are complete over $I_s$ on a Scc}{
    \nl       \If(\Comment{We do not know where we are}){$q = \bot$}{
    \nl          $(T,q,r) \gets \textsc{Home}(T,r,h)$
              }
    \nl       $(q',r',X,Y,r_1) \gets
                 \textsc{Transfer}(q,r,\Delta,\Lambda,I_1,J\setminus I_1)$
              \Comment{Target next transition to learn/sample} \\
    \nl       \If{such a path cannot be found}{
    \nl           goto line \ref{backbone:Scc}
                 \Comment{Graph not connected, try connecting with $I_s$}
              }
    \nl       \If(\Comment{$\pi(X)$ is not enabled in $q',r'$}){%
                 $Y = \Omega$ of $\omega$}{
    \nl          $\Lambda(q',\pi(X),\omega) \gets \{r',X,Y,r_1\}$,
                 $\Delta(q',\pi(X),Y) \gets q'$
                 \label{backbone:Omega} \\
    \nl          $q \gets q'$, $r \gets r'$
                 \Comment{We continue learning from the same state}
              }
    \nl       \Else(\Comment{$ {(q',r')} - X/Y \to \mathbf{(q'',r_1)} \to
                  w/\xi \to (q''',r_2)$}){%
    \nl           Let $q'' = \Delta(q',\pi(X),\pi(Y))$
                  \Comment{$q''$ (and so $\Delta$) might be a
                     ``state under construction''} \\
    \nl           \If(\Comment{We are sampling new values of a known transition
                            with same abstract output}){%
                      $q''$ is fully defined}{%
    \nl               $\Lambda(q',\pi(X),\pi(Y)) \gets
                      \Lambda(q',\pi(X),\pi(Y)) \cup \{(r',X,Y,r_1)\}$,
                      \Comment{Note we should not have a different $Y'$ with
                            same $r'$ unless (W-)inconsistency} \\
    \nl               $q \gets \Delta(q',\pi(X),\pi(Y))$, $r \gets r_1$
                  }
    \nl           \Else(\Comment{Learn tail of transition from $q'$
                              on input $X$}){%
    \nl               for some $w \not\in dom (\pi_1(q''))$,
                      \label{backbone:learn-trans} \\
    \nl                  \qquad $(T,\xi,r_2) \gets \textsc{Apply}(T,r_1,w)$
                         \label{backbone:add-to-dictionary} \\
    \nl               $\Lambda(q',\pi(X),\pi(Y)) \gets
                      \Lambda(q',\pi(X),\pi(Y)) \cup \{(r',X,Y,r_1)\}$ \\
    \nl               $\pi_1(q'')(w)) \gets \pi(\xi)$
                         \label{backbone:update-delta}
                         \Comment{This updates $\Delta$. $\pi_1$ is first element of couple} \\
    \nl               \If{$dom(\pi_1(q''))=W$}{%
    \nl                  $Q \gets Q \cup \{(\pi_1(q''),\rho_w(r_1,\epsilon)\}$\\
    \nl                  $A \gets \textsc{UpdateAccess}(T, q'',
                                                        \rho_w(r_1,\epsilon))$
                         \Comment{We record the shortest access
                                  to $(q'',r_1)$ from $(q,r)$ or the
                                  last homing in $T$} \\
    \nl                  $q \gets \Delta^*(q',\pi(X.w),\pi(Y.\xi))$
                         {if defined else}
                         $q \gets \bot $ \\
    \nl                  $r \gets r_2$}
    \nl              \Else{%
    \nl                  $q \gets \bot $}
                  }
    \nl           \If(\Comment{$\Delta^-$ trims the $\Delta$ graph from $h/a$ by cutting transitions that cannot be accessed with $A$ and $\Lambda$}){$\Delta^-(\Delta,\Lambda,h,a,A)$
                     (where $h/a$ was latest homing in $T$) is defined and
                      contains a complete strongly connected component (Scc)}{%
    \nl              $J \gets I_s$
                     \label{backbone:Scc}
                  }
              }
         }
    \nl \Return{$T, Q, \Delta, \Lambda$}
        \label{backbone:Exit}
    }
    \caption{Backbone procedure with guards on registers}
    \label{alg:backbone}
\end{algorithm}

\setcounter{AlgoLine}{0}

\SetKwFunction{Transfer}{\textrm{\textsc{Transfer}}}

\begin{algorithm}
    \nl \Fn(\Comment{We look for deterministic transfer, either through
                     unguarded transitions, or when registers and input
                     determine known transitions}){%
            \Transfer{$q, r, \Delta, \Lambda, I_1, I_2$}}{%
    \nl     \Comment{To reach a given input configuration $r'$, we record
                     the list of parameter values that can be set by unguarded
                     transitions while building the path, and can adapt the
                     parameter values at the end to set values to $r'$} \\
    \nl     \Comment{First we look for a short transfer with a bounded search
                     ($k \geq 0$ is the bound, tailorable), and if that fails,
                     we resort to a path from re-homing}
            \label{transfer:shortest-alpha} \\
    \nl     Find short(-est) $\alpha \in (\mathcal{I} \times O)^k$ and
            $X \in I_1$ with $\Delta^*(q,\pi(\alpha))=q'$ and \\
            $\rho_w(r,\alpha)=\pi_2(q')=\rho_w(r',\epsilon)$
            s.t. $\pi_1(\Delta(q',\pi(X),*))$ is partial,
            \Comment{First try to learn new input from a state, in its
                     reference configuration} \\
    \nl     \Comment{``partial'' means either not defined at all, or
                     there is an output whose characterisation is partial;
                     if $\Omega$ was the output for $\pi(X)$ it is not
                     partial regardless of $r'$ and $X$; if it was
                     $\omega$ for a given $r'$ and $X$, we can still look
                     for a different $r'$ or $X$.} \\
    \nl     or $\exists r' \neq \pi_2(q'), Y \neq Y'$
            s.t. $\rho_g(r,\alpha)=\rho_g(r',\epsilon)$ and \\
            $\{(\pi_2(q'),X,Y,*),(r',X,Y',*)\} \subset
            \Lambda(q',\pi(X),\pi(Y))$ and \\
            $\pi_1(\Delta(q',\pi(X),\pi(Y')))$ is partial
            \Comment{or a guarded transition} \\
    \nl     \If(\Comment{(otherwise) no transition to learn in current Scc}){%
                previous fails}{%
    \nl         Find $\alpha$ and $X \in I_2$ s.t.
                $\Lambda(q',\pi(X),*)$ does not contain $(*,X,*,*)$
                \Comment{transfer to a transition to be sampled} \\
    \nl         $q'=\Delta^*(q,\pi(\alpha))$, $r'=\rho(r,\alpha)$
            }
    \nl     \If(\Comment{bounded search failed, resort to homing}){%
                all previous fails}{%
    \nl         $(T,a,r) \gets \textsc{Apply}(T,r,h)$ and
                update $\Lambda$ on the way (if transitions are in $Q$) \\
    \nl         \If{$\Delta^-(\Delta,\Lambda,h,\pi(a),A)$ no longer contains
                    unsampled states and transitions}{%
    \nl             \Return{no path found}
                    \Comment{all transitions in reachable Scc already sampled
                             on $I_2$, transfer fails}
                }
    \nl         Look by BFS for shortest sequence
                $\alpha$ in $\Delta^-(\Delta,\Lambda,h,\pi(a),A)$,
                pick $A(\pi(a))(q',r')=\alpha$ and $X$ s.t.
                $\pi_1(\Delta(q',\pi(X),*))$ is partial or has guarded
                transition to partial state, or failing that has a transition
                to be sampled
            }
            \Comment{Here $q', r', \alpha, X$ are defined. If $\Delta$ was
                     partial in $q$, then $\alpha=\epsilon$, $q'=q$} \\
    \nl     $(T,\beta,r') \gets \textsc{Apply}(T,r,\alpha)$ and
            update $\Lambda$ on the way \\
    \nl     \If(\Comment{Transfer stopped prematurely on differing output}){%
                $\beta \neq \pi_o(\alpha)$}{%
    \nl         let $\beta=\beta'.o'$,
                $\alpha=\alpha'.X.\alpha''/\beta'.o.\beta''$
                s.t. $\pi_o(\alpha')=\beta', o' \neq o$ \\
    \nl         $Y \gets o'$, $r_1 \gets r', r' \gets
                   \rho(r,\alpha'), q' \gets \Delta^*(q,\alpha')$
                \Comment{We found a new guarded transition} \\
    \nl         \Comment{If h or W were not correct, we should handle
                         inconsistency if after $\alpha'$ we have same input
                         configuration $r$ for $o$ and $o'$}
            }
    \nl     \Else(\Comment{$(q,r) - \alpha(')/\beta(') \to
               \mathbf{(q',r')} - X/Y \to (q'',r_1) \to w/\xi \to (q''',r_2)$}){%
    \nl          $(T,Y,r_1) \gets \textsc{Apply}(T,r',X)$
            }
    \nl     $A \gets \textsc{UpdateAccess}(T,q',r')$
            \Comment{We record the shortest access to $(q',r')$ from $(q,r)$
                     or the last homing in $T$} \\
    \nl     \Return $(q',r',X,Y,r_1)$
        }
    \caption{Transferring from $q, r$ to next transition to learn}
    \label{func:transfer}
\end{algorithm}

\setcounter{AlgoLine}{0}


\begin{algorithm}
    \SetAlgoLined
    \nl \KwIn{$I_1 \subset I_2 \subset I_s \subset \mathcal{I}$,
            $W \subset I_1^*$, $R_w \subset R_g \subset R$}
    \nl     $h \in I_1^+$, s.t. $\rho_g(\bot,h)$ has a value
            for each register in $R_g$ \\
    \hspace*{-0.85em}
    \nl     \Init{$T \gets \epsilon$} \Comment{$T$ is the learning trace}\\
    \hspace*{-0.85em}
    \nl     \label{falg:repeat}
            \Repeat{no counterexample found}{%
    \nl         $Q, \Delta, \Lambda \gets \emptyset$, $I_o \gets I_1$ \\
    \hspace*{-0.85em}
    \nl         \Repeat{\textsc{Backbone} terminates with
                        no inconsistency}{%
    \nl             $T,Q,\Delta,\Lambda \gets$
                    \textsc{Backbone}($T,I_1,I_o,h,W,Q,\Delta,\Lambda$) \\
    \hspace*{-0.85em}
    \nl         handle inconsistencies on the way to update $h,W,I_1$
                \Comment{if h \& W not trustable} \\
    \hspace*{-0.85em}
    \nl         \For{$I_o$ incrementally ranging from $I_1$
                     to $I_1 \cup R_w(I_2)$}{%
    \nl              \Comment{Sampling on $R_w(I_2)$ first to have
                              the smallest set of register configurations
                              to compare, hence smallest number of redundant
                              states}

    \nl             $T,Q,\Delta,\Lambda \gets$
                    \textsc{Backbone} ($T,I_1,I_o,h,W,Q,\Delta,\Lambda$)
                }
    \nl         $I_o \gets I_1 \cup R_g(I_2)$ \\
    \hspace*{-0.85em}
    \nl         $T,Q,\Delta,\Lambda \gets$
                \textsc{Backbone} ($T,I_1,I_o,h,W,Q,\Delta,\Lambda$) \\
    \hspace*{-0.85em}
    \nl         \For{$I_o$ incrementally ranging from $I_1$
                     to $I_1 \cup R_w(I_2)$ then ${} \cup R_g(I_2)$}{%
    \nl              $(T,\mathit{CE}) \gets
                     \textsc{GetNFSMCounterExample}
                        (T,Q,\Delta,\Lambda,\mathit{SUL},I_o)$
                     \Comment{Ask for a CE using only inputs from $I_o$}  \\
    \hspace*{-0.85em}
    \nl              \If{CE found}{%
    \nl                  $(W,I_1,I_s,Q,\Delta,\Lambda) \gets$
                         \textsc{ProcessCounterexample}
                         \Comment{If W changed, this resets $Q,\Delta,\Lambda$}\\
                         \Comment{At this point, $I_s$ would not be changed} \\
    \hspace*{-0.85em}
    \nl                  continue to start of repeat \textsc{Backbone} loop
                     }
                }
    \nl         $I_o \gets I_s$
                \Comment{If $R_g$ is correct, $I_s$ will not change NFSM,
                         just feed \textsc{generalise}}
            }
    \nl     $(Q,\Delta,\Lambda) \gets
            \textsc{reduceFSM} (Q,\Delta,\Lambda)$
            \Comment{Reduce, not Minimize, as there is no unique minimum} \\
    \hspace*{-0.85em}
    \nl     \Repeat{$\neg$ (CE is a data CE)}{%
    \nl         $M \gets$
                \textsc{generalise}($T,h,Q,I,O,P_I,P_O,\Delta,\Lambda$) \\
    \hspace*{-0.85em}
    \nl         $(T,\mathit{CE}) \gets
                \textsc{GetCounterExample}(M,\mathit{SUL})$
            }
    \nl     \If{CE found}{%
    \nl         $(W,I_1,I_s,Q,\Delta,\Lambda) \gets$
                \textsc{ProcessCounterexample}
                \Comment{If W modified, $Q,\Delta,\Lambda$ are reset}
            }
        }
    \nl \Return{$M$}
    \caption{Main ehW algorithm}
    \label{alg:full-algo}
\end{algorithm}

\end{document}